# Computer-mediated therapies for stroke rehabilitation: a systematic review and meta-Analysis.

Stanley Mugisha PhD[1]. Mirko Job PhD[2]. Matteo Zoppi PhD[1], Marco Testa PhD[2], Rezia Molfino PhD[1].

1. Department of Mechanical, Energy, Management and Transport Engineering (DIME), University of Genoa, Italy

2. Department of Neuroscience, Rehabilitation, Ophthalmology, Genetics, Maternal, and Child Health, University of Genoa, Campus of Savona, Italy.

Details.

Stanley Mugisha.

Stanley.mugisha@edu.unige.it

+39 347 635 4760

Mirko Job

Mirko.job@edu.unige.it

+39 349 519 9788

Matteo Zoppi

Matteo.zoppi@unige.it

+39 0103 532 837

Marco Testa

Marko.testa@unige.it

+39 019 219 45701

Rezia Molfino

Rezia.molfino@unige.it

+39 010 33 52842

## Abstract

OBJECTIVE: To evaluate the efficacy of different forms of virtual reality (VR) treatments as either immersive virtual reality (IVR) or non-immersive virtual reality (NIVR) in comparison to conventional therapy (CT) in improving physical and psychological status among stroke patients.

METHODS: The literature search was conducted on seven databases: ACM Digital Library, Medline (via PubMed), Cochrane, IEEE Xplore, Web of Science, Scopus, and science direct. The effect sizes of the main outcomes were calculated using Cohen's d. Pooled results were used to present an overall estimate of the treatment effect using a random-effects model.

RESULTS: A total of 22 randomized controlled trials were evaluated. 3 trials demonstrated that immersive virtual reality improved upper limb activity, function and activity of daily life in a way comparable to CT. 18 trials showed that NIVR had similar benefits to CT for upper limb activity and function, balance and mobility, activities of daily living and participation. A comparison between the different forms of VR showed that IVR may be more beneficial than NIVR for upper limb training and activities of daily life.

CONCLUSIONS: This study found out that IVR therapies may be more effective than NIVR but not CT to improve upper limb activity, function, and daily life activities. However, there is no evidence of the durability of IVR treatment. More research involving studies with larger samples is needed to assess the long-term effects and promising benefits of immersive virtual reality technology.

## Introduction.

Stroke has been described as one of the significant causes of death and disability globally, representing a severe problem for public health with a significant prevalence in men and women of all ages [1,2]. Recovery is always incomplete, and most survivors are left with motor, sensory and cognitive impairments with a consequent increase in the burden of health care expenses during adulthood [3].

Owing to the rising number of neurologically impaired survivors, several computer-mediated programs for stroke rehabilitation have recently been developed to help patients regain their ability to live independently. In particular, the advancement of digital technology has favored the assertation of virtual reality (VR) as an accessible solution to give patients feedback on their performance, meaningful goals, and a personalized experience to support motor learning [4,5]. Training can be gamified through various applications, making the rehabilitation process fun and enjoyable. VR is a real-time, computer-based, interactive, multisensory simulation environment that enables users to engage in activities inside environments that resemble real-world artifacts and events to varying degrees [6–8].

Depending on the quantity of visual sensory channels engaged in simulation, VR can be categorized as either immersive or non-immersive. Immersive VR (IVR) replaces the user's real-world environment with a simulated [9,10]. Users get the sense of being transported into three-dimensional interactive worlds through 360° immersion in an alternate reality such as a head-mounted display (HMD) or video capture systems such as IREX, which enables them to participate in various activities in imaginary environments[11–13].

In non-immersive VR (NIVR), the user mainly interacts with virtual objects displayed either in a 2D or 3D environment that can be directly manipulated on a conventional graphics workstation using a keyboard and a mouse. As in IVR, animation and simulation are interactively controlled to the user's direct manipulation with some NIVR systems and allow individuals to see their avatars reflected on the screen [14–17].

Non-immersive systems are characterized by a lower level of immersive features (e.g., scene changes with head movements), which could play a role in supporting the feeling of presence and its therapeutic benefits [13]. For instance, it has already been pointed out how immersion in the virtual simulation plays a pivotal role in pain management by inducing relaxation.

Previous reviews have studied the use of VR in the rehabilitation of stroke [18–21]. However, none of these works differentiated their results according to the VR modalities. Since the type of VR seems to influence the rehabilitation outcomes differently depending on the level of immersion [18,19], it is vital to unravel which form yields the best treatment effects for motor rehabilitation and other outcomes important to people with stroke such as quality of life and participation.

Hence, by differentiating between IVR and NIVR systems, this systematic review and meta-analysis of randomized control trials aimed to evaluate the efficacy of VR treatments in improving physical and psychological status in patients with stroke compared to conventional therapy (CT).

## Methods

### Registration Number.

The review protocol and inclusion criteria were pre-specified and registered on the National Health Service Prospero Database under the registration number: CRD42019134806. This systematic review followed the Preferred Reporting Items for Systematic Reviews and Meta-Analyses guidelines.

Electronic searches.

The literature search was conducted on the following databases: MEDLINE (via PubMed), Scopus, Web of Science, ACM Digital Library, and IEEE Xplore. Additional articles were retrieved by scanning the reference lists of those studies that passed the "full-text" screening stage. The search strategy was organized by extracting a set of keywords from three primary groups representing:(i) technology, (ii) rehabilitation, and (iii) pathology-related semantic fields. Terms were connected using the OR boolean operator for within-group connections and the AND boolean operator between-group relations. Both free and MeSH terms were used to build the final search string, reported in detail in Appendix I. Results obtained from the database research were filtered to include only those published after 2015, chosen as the lower temporal limit because Immersive Virtual reality is relatively a new technology, and its use in neurological rehabilitation is still in the early stages. Other works were retrieved by scanning the reference lists of all included studies.

Eligibility Criteria

To be included, articles had to meet the following eligibility criteria. (i) a randomized controlled trial that (ii) considered subjects more than 18 years old, (ii) affected by neurological disorders, (iii) and compared computer-mediated treatments against conventional therapy (iv) for the upper limbs or lower limb motor functions (v) or postural control. Moreover, only full-text articles written in English were included in the screening process. We excluded those articles that: (i).compared one or more different types of computer-mediated reality-based treatments without an alternative control group; (ii) involved high-cost devices such as treadmills, CAVE, and any form of a robotic manipulator.
(iv) Other neurological conditions apart from a stroke.

Study selection.

In May 2020, we started a comprehensive systematic search. Duplicates across databases were removed, and the remaining studies were screened for titles, abstracts, and descriptors by the two reviewers independently (MS, MJ) to assess whether they met the predefined inclusion criteria. Controversies between the reviewers concerning the eligibility were resolved in a consensus meeting. After reaching an agreement, full texts of potentially eligible studies were retrieved and further assessed against the inclusion criteria, and reasons for excluding the studies were documented.

Risk of Bias

We used the Critical Appraisal Skills Programme for randomized controlled trials [22] to assess the risk of bias in the included trials. It is 11 questions checklist with three sections that assess the following items: validity of study results, what the results are, and how they are helpful locally. Each question in each subsection required a positive('yes'), neutral (Can't tell), or negative answer('No').
The articles were classified as low risk, moderate risk, or high risk of bias according to the number of items that received a negative appraisal. Those articles in which all checklist items were appraised positively were considered low risk of bias works. Articles in which one or two of the checklist items were appraised negatively were considered a moderate risk of bias. Articles in which three or more items were appraised negatively were considered high risk of bias works.
The two authors independently assessed the quality of the work, and any disagreements that arose during the process were resolved through discussion.

Data Extraction.

Two review authors (MS and MJ) independently extracted data into a
a custom data table and the following data were chosen to be extracted from each study:
(i) citation details, (i) Population characteristics; (ii) inclusion and exclusion criteria (ii) Type of intervention (iii) Technology Used (iv) outcome measures; (v) main results. One author (M.S) extracted data, and another (M.J)checked for accuracy. For studies where values were provided in an unconventional format (i.e., medians [interquartile range], or means [minimum-maximum range][23–25]) the sample mean and standard deviation were estimated as described in Wan et al. 2014 [26]. From this, the effect sizes of the main outcomes were calculated.

Data synthesis

Two review authors (MS and MJ) independently classified outcome measures in terms of the domain assessed ((i)Upper limb activity and function (ii) lower limb activity and function, (iii) balance, (iv) activity of daily life, and (v) adverse events). When more than one outcome measure for the same domain was presented in a study, the most frequently used across studies was considered in the analysis. The standardized mean differences (SMD) were calculated for continuous outcomes and Cochrane's Review Manager 5 (Review Manager 2014) software was used for all analyses. The effect of the intervention was measured using cohen's d on the primary outcome of each study. In trials with three-armed interventions [27], the VR therapy was compared to conventional interventions.

Meta-analysis

Pooled results were used to present an overall estimate of the treatment effect using a random-effects model in the analysis across studies. Heterogeneity was assessed through the $I^2$ statistic [28]. The level of heterogeneity was considered substantial if the $I^2$ statistic was greater than 50%. A meta-analysis was considered not appropriate where the level of heterogeneity was substantial or when only one study was identified for the desired outcome. In this case, a narrative summary of the results was given. Where data pooling was decided, forest plots were provided along with a description of the results. When applicable, a sensitivity analysis was performed, including only studies at low risk of bias, and the results were compared to the primary analysis, including all the trials.

Results:

Search Results.

The study inclusion workflow is displayed in detail in Figure 1. We identified a total of 1573 possible record from the database research (ACM: n=18,Pubmed: n=1382, Cochrane: n=17, IEEE: n=29, Science direct: n=60, Scopus: n=38, Web of Science: n=29). Additional 151 Records were identified through secondary sources. After removing records duplicates, 1680 studies were evaluated for title and abstract, resulting in 95 full-text articles assessed against the previously defined eligibility criteria. At the end of this process, 22 full- text articles were included in the present review (Adie et al. 2017; Allen et al. 2017; Aşkın et al. 2018; Choi, Shin, and Bang 2019; Henrique et al. 2019; Huang and Chen 2020; Ikbali Afsar et al. 2018; In, Lee, and Song 2016; Kiper et al. 2018; K. H. Kong et al. 2016; Llorens et al. 2015; McNulty et al. 2015; Mekbib et al. 2021; Ögün et al. 2019; Pedreira da Fonseca et al. 2017; Saposnik et al. 2016; Schuster-Amft et al. 2018; Shin et al. 2016; Shin, Bog Park, and Ho Jang 2015; Da Silva Ribeiro et al. 2015; Bin Song et al. 2015; Zondervan et al. 2016. And these 2 studies [29,30] were excluded from the meta-analysis.

40 studies did not meet the inclusion criteria because they were published before 2015. 8 studies were excluded from this review because the VR intervention methods used expensive devices such as treadmills and robotic manipulators or actuated devices [31–36,37]. 3 studies were excluded because they compared virtual reality with no intervention [38–40].

Risk of bias.

The critical quality appraisal of the 22 included trials is presented in Table 1. Relying on the CASP guidelines [22] for a randomized trial, we identified 2 of the included studies to be at high risk of bias [42,47], 3 at low risk of bias [17,46,54] and 17 at moderate risk of bias [8,23,25,27,29,30,41,43–45,48–53,55]. In 2 trials [42,47], the authors did not report the information about the randomization method of patients. In [48], the treatment between groups was not the same. The experimental group received 30 extra minutes of VR training for 20 sessions making up 10 hours. 7 studies reported a large treatment effect of the Intervention measured using Cohen's d [17,42,45,47–49,52,54], while the rest had a medium to small effect size. It is also important to consider that small sample sizes characterized most studies. A total of 13 studies Aşkın et al., 2018; Bin Song et al., 2015; Choi et al., 2019; Da Silva Ribeiro et al., 2015; Huang & Chen, 2020; In et al., 2016; H.-C. Lee et al., 2017; Llorens et al., 2015; McNulty et al., 2015; Mekbib et al., 2021; Pedreira da Fonseca et al., 2017; Saposnik et al., 2016; Zondervan et al., 2016) had a small sample population of about 10 to 16 participants in the experimental group, resulting in broader confidence intervals and therefore were regarded to be of high risk of bias. Due to the nature of the study, blinding of the patients, physiotherapists, and assessors who supervised the treatments was not possible. Therefore it was not considered as a possible risk of bias.

Population Characteristics.

Of all 22 trials, 7 trials had a sample size of more than 50 participants [8,17,27,41,45,50,53]. And 6 trials had less than 25 participants [47,49,51,52,55,56].

*Inclusion and Exclusion criteria*

Most studies specified the inclusion and exclusion criteria apart from two where the exclusion criteria were not given [42,49]. Most Participants in the included studies appeared to be relatively young, with mean ages ranging from 29 to 75 years in all studies. Studies omitted medically ill participants, such as specified by the presence of a disease in which exercise was contraindicated. Participants were included if they were cognitively intact, as defined by cut-off scores on the MMSE. Medically unstable participants were excluded, as defined by having a disease in which exercise was contraindicated, severe visual disorders [46,51,55] and neurodegenerative disorders[23].

*Interventions.*

In all retrieved trials, the active control group performed similar exercises as those proposed in the VR intervention. For IVR, 3 studies implemented an equally matched dose of CT [17,53,55], while 1 trial consisted of a combination of VR with CT [54].

For NIVR interventions, 7 trials had the intervention group that performed only VR exercises [8,27,29,41,42,44,46], 7 studies had a VR intervention group performing a combination of VR exercises augmenting CT [23,25,30,45,48–51].

One trial consisted of three-armed interventions including a VR group, an active control group performing similar exercises within a conventional physiotherapy setting, and a passive control group [43].

In all trials, therapy sessions lasted between 45 minutes and 2 hours per day, with a minimum duration of therapy lasting for 16 hours and a maximum of 40 hours.

*Technology.*

Participants in the IVR group received training using immersive devices such as head-mounted displays [17,47,52], a leap motion controller [17,52], and Htc Vive controllers in [47]. While For NIVR interventions, participants mainly used the Microsoft Xbox Kinect connected to a display monitor [23,42,46,48], Nintendo Wii gaming system [8,27,29,30,41,44,45], motion tracking sensors [25,50,51], and data gloves [53–55].

A detailed overview of the IVR and NIVR interventions is provided in Table 4 and Table 5, respectively.

Outcomes.

Due to various intervention approaches, a wide range of outcome measures was retrieved. The outcome measures were collected at the baseline and soon after the intervention. In all the trials, a post-intervention follow-up assessment was done in only 7. Of these, 5 had a post-intervention follow-up of less than 3 months [8,27,45,54,55], and 2 had more than 6 months [30,41]. An overview of all outcome measures for each predefined outcome category and results for the primary outcome in the included studies can be found in tables 6 and 7 for IVR and NIVR respectively.

Immersive Virtual Reality vs CT *(Short term Effects)*

*Upper limb activity and function*

Upper limb function was accessed using FM as the outcome measure. In 3 trials with 106 participants [17,47,52], improvements were reported in both training groups post-intervention. A meta-analysis revealed significant inter group differences with greater improvements in the IVR group compared to CT (smd:1.37, 95% CI [0.80, 1.93], $I^2$=35%, P<0.00001). A forest plot is shown in figure 2.

A sensitivity analysis of results without [47] considered to be a high risk of bias revealed similar results(smd:1.57, 95% CI [1.03, 2.11], $I^2$=15%, P=0.2).

Upper limb activity was assessed by BBT as the outcome measure. Results from one study [47] with 18 participants reported improvements in both groups. IVR registered a higher improvement than CT. However, the difference between the groups was not significant IVR:(0.08±0.14), CT(0.04±0.06), (smd= 0.35, 95% CI [-0.58, 1.29], p>0.05). This was a low-quality study with large confidence intervals.

*Activities of daily life (ADL)*

Three trials [17,47,52] investigated the effects of IVR on ADL. A meta-analysis indicated greater improvement in the IVR group with a moderate effect size (smd = 0.54 ,95% [CI 0.15 to 0.93], $I^2$ = 0%, p=0.007). A forest plot is shown in figure 3.

A sensitivity analysis without [47] considered high risk of bias revealed similar results (smd = 0.61 ,95% [CI 0.18 to 1.04], $I^2$ = 0%, p=0.005).

*Adverse events.*

In all studies, the occurrence of adverse events was not reported.

Non-Immersive VR Vs CT (Short Term Effects)

*Upper limb activity and function*

Nine trials reported results on upper limb function using FM as the outcome measure [23,25,44,46,48,50,54,57]. Significant improvements were reported in both groups post-intervention. A meta-analysis was not performed because of high heterogeneity ($I^2$ = 82%) in the data. According to results from individual studies, NIVR reported greater improvements than CT.

Four trials with a total population of 172 participants reported greater improvement in NIVR compared to CT with a large effect size , Aşkın et al. 2018 [23],(NIVR:41.25±9.0, CT:35.00±10.0, smd=0.64), Ikbali Afsar et al. 2018 [48](NIVR:18.74±7.67, CT:13.94±6.58,

smd=0.65), Shin et al. 2016 [54] [NIVR:4.9±1.0, CT:1.4± 0.8, smd=3.71] and Henrique et al. 2019 [46] [NIVR: 14.69±0.67, CT:9.07±1.34, smd = 5.22],

Four studies with a total of 265 participants reported no significant differences between the groups. Kong et al. 2016 [27] (NIVR:32.8±18.2, CT:29.2±17.25, smd=0.2), Kiper et al. 2018 [50] (NIVR:47.1±15.74 , CT:46.29±17.25, smd=0.09) Shin, Bog Park, and Ho Jang 2015[25], (NIVR:38.5±11.7, CT:33.87±17.64, SMD=0.3] and Da Silva Ribeiro et al. 2015[44] (NIVR: 38.7±19.6 CT: 44.7±14.2, smd=-0.34).

In one study [30], with 41 participants, we could not get data in a suitable format for analysis. However, they reported no significant difference between groups.

Three trials with 109 participants measured upper limb activity using BBT as the outcome measure [23,53,55]. A meta-analysis revealed no significant difference between the groups. The overall effect size was small (smd = 0.19, 95% CI [-0.19, 0.57], $I^2 = 0$%, P=0.33) as shown by a forest plot in figure 4.

*lower limb activity and function*
Three trials with a total of 92 participants used TUG to assess mobility, [42,45,49]. Improvements were reported in both groups post-intervention. A meta-analysis done revealed no significant difference between the groups (smd = 0.33, 95% CI [-0.08, 0.75], $I^2 = 0$% , P=0.11) as shown by a forest plot in figure 5. A sensitivity analysis that excluded results from one study [42] deemed to be a high risk of bias revealed similar results (smd = 0.34, 95% CI [-0.12, 0.81], $I^2 = 0$%, P=0.15).

In 3 trials with 65 participants, the 10mw Test was used as the outcome measure [42,45,51], A meta-analysis revealed no significant difference between the groups (smd = -0.06, 95% CI [-0.60, 0.48] ,$I^2 = 0$% P=0.82). A forest plot is shown in figure 6. A sensitivity analysis which excluded results from [42] (smd = -0.22, 95% CI [-1.00, 0.56] ,$I^2 = 42$% P=0.58) revealed similar results.

*Balance.*
Four trials with 123 participants used BBS as the outcome measure for balance [45,46,49,51]. The VR group had greater improvements with a moderate overall effect size. (smd:0.46, 95% CI [-0.01, 0.93], $I^2$=37%, P=0.06). However, the results were not statistically significant. A forest plot is shown in figure 7.

*Quality Of Life and participation*
Eight trials with 686 participants measured Quality Of Life using various outcome measures. [8,25,27,41,43–45,50]. A meta-analysis showed no significant difference between groups with (smd = 0.04 95%, CI[-0.11, 0.19], $I^2$= 0%, P=0.79) . A forest plot is shown in figure 8.

*Activities of daily life.*
In nine trials [8,27,30,45,48,50,53–55] with 550 participants and several outcome measures**.** A meta-analysis revealed a significant difference between the groups but the overall effect size was low. However, the statistical heterogeneity was moderate. (smd = 0.25, 95% CI, [0.02 to 0.49], $I^2$=43.0%, p=0.04). A forest plot is shown in figure 9.

Non-Immersive Virtual Reality Vs CT(Long term Effects).
*Upper limb Function and Activity*
Two trials, studied the long-term effects of VR on upper limb function using FM as the outcome measure. A meta-analysis was not done due to high statistical heterogeneity in the data. However, in [27] with a population of 64 participants, after 15 weeks follow up, greater improvements were reported in the VR group but there was no significant difference between the treatment groups. The overall effect size was small. (VR:40.4 ± 20.7, CT:34.5± 19.5, smd=0.29, 95% CI[-0.20,0.78]). While in [54], with a population of 23 participants, after 1-month follow-up, VR significantly improved with a large effect size (VR:5.3±1.1, CT=1.3±0.8, smd=3.87, 95% CI[2.38,5.36]).

In two studies [53], [8] with 153 participants, upper limb activity was measured by BBT. After 2 months [53] and 4 weeks [8] follow up assessment, there were no significant differences between the groups. (smd = -0.06 95% CI [-0.38, 0.26], $I^2$= 0% P = 0.78). A forest plot is shown in figure 10.

*Lower limb activity.*
Only one trial [45] with a total of 47 participants reported long-term effects on lower limb activity. Three months after intervention using TUG, greater improvements were reported in the VR group. However, there was no significant inter-group difference (NIVR: -23.52±10.96, CT:-28.67±18.73, smd = 0.34, 95% CI [-0.24, 0.92]).

*Balance*
In One trial [45] which used BBS as the outcome measure, a follow-up assessment conducted 3 months post-intervention with 137 participants reported no significant difference between the two intervention groups. (VR: 46.31±5.8 , CT 45±5.06  smd= 0.23,  95% CI [-0.34 to 0.81]).

*Quality Of Life and participation*
Four trials [8,27,41,45] with 406 participants, conducted a post intervention follow-up on the quality of life. Improvements were noted in both groups without significant intergroup difference (smd=0.04, 95% CI[ -0.15 to 0.24], $I^2$=0.0%, p= 0.67). A forest plot is shown in figure 11.

*Activities of Daily Living*
6 trials [8,27,30,41,45,53] with a total of 493 participants, reported a follow-up post-intervention. A meta-analysis of results revealed no significant difference between the groups. (smd, = 0.00 95% CI [-0.22, 0.23], $I^2$=28.0%, p= 0.97). A forest plot is shown in figure 12.

*Adverse events.*
Eight trials monitored adverse events [8,23,27,29,30,45,58,59]. However, no serious adverse event related to the treatments was reported.

## Discussion.

This work investigated the effectiveness of IVR and NIVR compared to conventional therapy for stroke rehabilitation based on 22 included trials.

In general, the statistical analysis revealed that both VR interventions positively affected patients' functionality in a comparable way to CT.

As a comparison of NIVR to CT, this meta-analysis discovered positive improvement in treatments in favor of VR with small to medium effect sizes but with no significant difference between the different techniques. Higher values of effect sizes in favor of the NIVR indicated that patients had improvements in upper limb function and activity, mobility, balance, and ADL. However, their level of independence, which is the aim of rehabilitation strategies did not improve as their counterparts who received CT. The magnitude of this effect was comparable to that observed in previous systematic reviews. They concluded that using VR-based therapy systems enhanced upper limb function, quality of life [18,20], mobility, and balance [21] in people with stroke but not significantly greater than CT. There is an indication that VR may alleviate upper limb motor impairments and encourage motor activities and societal participation among stroke survivors.

More improvements in VR could be due to several crucial factors. The first may be the ability to provide therapists with various training programs. The therapist may select different treatment modes and construct an individualized training program that adapts the intensity and difficulty level of the training to the patient's current motor status [60]. The potential of VR to scale difficulty levels and give adequate rewards to users in the context of gaming and level progression is vital to the implementation of effective VR training systems. Secondly, VR systems incorporate task-specific workouts in addition to an appropriate level of exercise intensity and repetition, from which patients can benefit. The majority of the included trials in our review comprised graded training regimens to induce optimal neural plasticity and continuous active engagement, both of which are essential for successful motor recovery after a stroke [61]. A recent review found that custom-built VR systems had a more significant effect on the recovery of upper limb extremity function and activity than using CT [62]. Customized VR systems are usually constructed based on a valid hypothesis to provide an effective rehabilitation regimen beneficial to patients [19,61,63]. Therefore, the ability to configure a VR therapy system with multiple training alternatives may be critical for upper-limb rehabilitation of motor deficits following stroke. Another reason could be the numerous types of sensory feedback like visual and aural present in VR therapeutic systems, which make the training more enjoyable. The use of virtual reality (VR) can assist in the creation of environments in which more repeated actions are done in a playful context, hence enhancing motivation and adherence to therapy. The majority of the selected studies have taken into account this essential factor, either through participation scales or questionnaires assessing the level of motivation.

By differentiating the effects obtained by immersive and non-immersive solutions, our evidence suggests that IVR may be more beneficial than NIVR for upper limb function and daily life activities. However, it is important to note that the results of IVR intervention are based on short-term effects and few studies with a limited number of participants. We observed that the effect size values of the results from studies that used IVR were higher than NIVR.

A possible reason behind this phenomenon may be the missing depth cues in the 2D environments that characterize NIVR applications [64].

Piggott et al. (2016) observed in their study that subjects using 2D VR systems such as cyber gloves tend to decrease their wrist extension [65]. This could be resolved by using a head-mounted display (HMD). On the contrary, IVR systems take into account depth cues [17], which are responsible for accelerated cortical reorganization[66], thus allowing the central nervous system to control the position and orientation of body segments and adapt to the simulated environment.

The effect of different neurological characteristics on VR rehabilitation outcomes also needs examination. Some studies suggest that hemorrhagic stroke may result in more severe cognitive, motor, and functional impairment than ischemic stroke [68,69]. Future investigations would benefit from a comparison of these stroke types to test the impact of both forms of VR.

The absence of adverse events related to the treatment suggests that NIVR can be considered a safe treatment. This data is consistent with results from a review by Laver et al. [18], who found little or no adverse events from VR treatment after stroke. This was because most studies took place in settings that applied extra safety measures such as supervision or walking harnesses in a laboratory or rehabilitation setting. It would also be important to note that IVR studies did not monitor occurrence of adverse events among participants, therefore it would important to examine the safety and psychological outcomes associated with the use of head mounted displays.

Limitations.

The presented results should be interpreted considering some limitations. High heterogeneity in the data made it difficult to perform a meta-analysis of results for some of the outcomes of interest. For example, a meta-analysis on the beneficial effects of NIVR on upper limb function using FM was not feasible.

There was a high diversity between the VR training scenarios, which made comparing the results across the studies difficult. Therefore, we could not make a firm conclusion on the benefits of this technology compared to CT, though all studies reported post-intervention improvements.

The IVR studies were single-center designs characterized by high dropout rates mainly due to medical reasons and compliance issues. [17] reported a high dropout rate in the CT group (22.5%) and (23%) in the IVR group. Our results should be validated on additional works based on larger populations.

Another limitation was the wide variety of outcome measures used in the included articles, which precluded the possibility to compare different evidence across studies, as trials used different versions of the same scale or different measuring units.

NIVR is already established and has many studies. On the contrary, the utilization of IVR systems for motor rehabilitation programs is still in the early stages. IVR is a relatively new technology and remains partially known, with a lot of the work limited to pain and phobias treatment [13,67]. There are few RCTs on the effectiveness of immersive virtual reality systems in stroke rehabilitation. Furthermore, we did not find studies that examine the long-term benefits of IVR, and therefore more trials are needed to validate the intensity of efficacy.

Additionally, most of the included trials had small sample sizes, which resulted in low certainty in the effect measures and low statistical power. We recommend larger studies in the future, with power calculations pointing to more than 25 participants per group.

## Conclusion

This study on the efficacy of virtual reality therapies applied to the rehabilitation of patients with stroke highlights the benefits of VR. However, evidence of the clinical effectiveness of the different forms of virtual reality (as either immersive or non-immersive) is scarce. Results from this review suggest that IVR therapies may be more effective than NIVR but not CT to improve upper limb activity, function, and daily life activities. The results of IVR intervention are based on short-term effects with small effect sizes. Therefore, there is no evidence that IVR treatment is long-lasting. NIVR provides the same benefits as CT for mobility, balance, quality of life, and daily life activities among patients with stroke. While the current literature evaluates VR as a viable alternative to conventional therapy in stroke rehabilitation, much attention is accorded to non-immersive solutions due to their wide use in clinical and research fields. By exploiting the increasing use and availability of IVR systems, additional controlled trials with larger sample sizes should be carried out in the future to reliably assess the long-term effects and promising benefits of this technology.

**Author contributions**
 MS and MJ contributed to the study design, data collection, data interpretation and drafted the manuscript. MT and MZ contributed to the study design, data interpretation, and manuscript revision. RM contributed to the study design and revision of the manuscript. All authors have revised and approved the manuscript.
**Declaration of interests**: None
**Declaration of Funding sources/sponsors**: None

1. Sims, N. R. & Muyderman, H. Mitochondria, oxidative metabolism and cell death in stroke. *Biochim. Biophys. Acta* **1802**, 80–91 (2010).
2. Feigin, V. L., Norrving, B. & Mensah, G. A. Global Burden of Stroke. *Circ. Res.* **120**, 439–448 (2017).
3. Levine, D. A. *et al.* Recent trends in cost-related medication nonadherence among stroke survivors in the United States. *Ann. Neurol.* **73**, 180–188 (2013).
4. Deutsch, J. E. E. & Westcott McCoy, S. Virtual Reality and Serious Games in Neurorehabilitation of Children and Adults: Prevention, Plasticity, and Participation. *Pediatr. Phys. Ther.* **29**, S23–S36 (2017).
5. Perrochon, A., Borel, B., Istrate, D., Compagnat, M. & Daviet, J.-C. Exercise-based games interventions at home in individuals with a neurological disease: A systematic review and meta-analysis. *Ann. Phys. Rehabil. Med.* **62**, 366–378 (2019).
6. Rizzo, A. "Skip" & Kim, G. J. A SWOT Analysis of the Field of Virtual Reality Rehabilitation and Therapy. *Presence* **14**, 119–146 (2005).
7. Henderson, A., Korner-Bitensky, N. & Levin, M. Virtual Reality in Stroke Rehabilitation: A Systematic Review of its Effectiveness for Upper Limb Motor Recovery. *Top. Stroke Rehabil.* **14**, 52–61 (2007).
8. Saposnik, G. *et al.* Efficacy and safety of non-immersive virtual reality exercising in stroke rehabilitation (EVREST): a randomised, multicentre, single-blind, controlled trial. *Lancet Neurol.* **15**, 1019–1027 (2016).
9. Kim, A., Darakjian, N. & Finley, J. M. Walking in fully immersive virtual environments: an evaluation of potential adverse effects in older adults and individuals with Parkinson's disease. *J. Neuroeng. Rehabil.* **14**, 16 (2017).
10. Luis, M. A. V. S., Atienza, R. O. & Luis, A. M. S. Immersive virtual reality as a supplement in the rehabilitation program of post-stroke patients. in *International Conference on Next Generation Mobile Applications, Services, and Technologies* 47–52 (IEEE Computer Society, 2016). doi:10.1109/NGMAST.2016.13.
11. Kyriakou, M., Pan, X. & Chrysanthou, Y. Interaction with virtual crowd in Immersive and semi-Immersive Virtual Reality systems. *Comput. Animat. Virtual Worlds* **28**, 1–12 (2017).
12. Rose, T., Nam, C. S. & Chen, K. B. Immersion of virtual reality for rehabilitation - Review. *Appl. Ergon.* **69**, 153–161 (2018).
13. Colloca, L. *et al.* Virtual reality: physiological and behavioral mechanisms to increase individual pain tolerance limits. *Pain* **161**, 2010–2021 (2020).
14. Turolla, A. *et al.* Virtual reality for the rehabilitation of the upper limb motor function after stroke: a prospective controlled trial. *J. Neuroeng. Rehabil.* **10**, 85 (2013).
15. Trevizan, I. L. *et al.* Efficacy of different interaction devices using non-immersive virtual tasks in individuals with Amyotrophic Lateral Sclerosis: a cross-sectional randomized trial. *BMC Neurol.* **18**, 209 (2018).
16. Mirelman, A. *et al.* Addition of a non-immersive virtual reality component to treadmill training to reduce fall risk in older adults (V-TIME): a randomised controlled trial. *Lancet* **388**, 1170–1182 (2016).
17. Ögün, M. N. *et al.* Effect of leap motion-based 3D immersive virtual reality usage on upper extremity function in ischemic stroke patients. *Arq. Neuropsiquiatr.* **77**, 681–688 (2019).
18. Laver, K. E. *et al.* Virtual reality for stroke rehabilitation ( Review ). *Cochrane database Syst. Rev.* **11**, CD008349 (2018).
19. Mekbib, D. B. *et al.* Virtual reality therapy for upper limb rehabilitation in patients with stroke: a meta-analysis of randomized clinical trials. *Brain Inj.* **34**, 456–465 (2020).
20. Saposnik, G. & Levin, M. Virtual reality in stroke rehabilitation: A meta-analysis and implications for clinicians. *Stroke* **42**, 1380–1386 (2011).
21. Corbetta, D., Imeri, F. & Gatti, R. Rehabilitation that incorporates virtual reality is more effective than standard rehabilitation for improving walking speed, balance and mobility after stroke: a systematic review. *J. Physiother.* **61**, 117–124 (2015).
22. CASP-UK. CASP Randomised Controlled Trial Checklist. *CASP checklists Randomised Control. Trial* 1–7 (2018) doi:http://media.wix.com/ugd/dded87_40b9ff0bf53840478331915a8ed8b2fb.pdf.
23. Aşkın, A. *et al.* Effects of Kinect-based virtual reality game training on upper extremity motor recovery in chronic stroke. *Somatosens. Mot. Res.* **35**, 25–32 (2018).
24. Schuster-Amft, C. *et al.* Using mixed methods to evaluate efficacy and user expectations of a virtual reality-based training system for upper-limb recovery in patients after stroke: a study protocol for a randomised controlled trial. *Trials* **15**, 350 (2014).
25. Shin, J. H., Bog Park, S. & Ho Jang, S. Effects of game-based virtual reality on health-related quality of life in chronic stroke patients: A randomized, controlled study. *Comput. Biol. Med.* **63**, 92–98 (2015).
26. Wan, X., Wang, W., Liu, J. & Tong, T. Estimating the sample mean and standard deviation from the sample size, median, range and/or interquartile range. *BMC Med. Res. Methodol.* **14**, 1–13 (2014).
27. Kong, K. H. *et al.* Efficacy of a virtual reality commercial gaming device in upper limb recovery after stroke: A randomized, controlled study. *Top. Stroke Rehabil.* **23**, 333–340 (2016).
28. Higgins JPT, Thomas J, Chandler J, Cumpston M, Li T, Page MJ, W. V. *Cochrane Handbook for Systematic Reviews of Interventions*. (Cochrane, 2019).
29. Pedreira da Fonseca, E., da Silva Ribeiro, N. M., Pinto, E. B., Ribeiro da Silva, N. M. & Pinto, E. B. Therapeutic Effect of Virtual Reality on Post-Stroke Patients: Randomized Clinical Trial. *J. Stroke Cerebrovasc. Dis.* **26**, 94–100 (2017).
30. McNulty, P. A. *et al.* The efficacy of Wii-based Movement therapy for upper limb rehabilitation in the chronic poststroke period: A randomized controlled trial. *Int. J. Stroke* **10**, 1253–1260 (2015).
31. Oh, Y.-B. *et al.* Efficacy of Virtual Reality Combined With Real Instrument Training for Patients With Stroke: A Randomized Controlled Trial. *Arch. Phys. Med. Rehabil.* **100**, 1400–1408 (2019).
32. de Melo, G. E. L. *et al.* Effect of virtual reality training on walking distance and physical fitness in individuals with Parkinson's disease. *NeuroRehabilitation* **42**, 473–480 (2018).
33. Calabro, R. S. *et al.* The role of virtual reality in improving motor performance as revealed by EEG: a randomized clinical trial. *J. Neuroeng. Rehabil.* **14**, 53 (2017).
34. Calabro, R. S. *et al.* Robotic gait training in multiple sclerosis rehabilitation: Can virtual reality make the difference? Findings from a randomized controlled trial. *J. Neurol. Sci.* **377**, 25–30 (2017).
35. Peruzzi, A., Zarbo, I. R., Cereatti, A., Della Croce, U. & Mirelman, A. An innovative training program based on virtual reality and treadmill: effects on gait of persons with multiple sclerosis. *Disabil. Rehabil.* **39**, 1557–1563 (2017).
36. Liao, Y.-Y. Y. *et al.* Virtual Reality-Based Training to Improve Obstacle-Crossing Performance and Dynamic Balance in Patients With Parkinson's Disease. *Neurorehabil. Neural Repair* **29**, 658–667 (2015).
37. Lee, S., Kim, Y. & Lee, B.-H. H. Effect of Virtual Reality-based Bilateral Upper Extremity Training on Upper Extremity Function after Stroke: A Randomized Controlled Clinical Trial. *Occup. Ther. Int.* **23**, 357–368 (2016).
38. Alves, M. L. M. *et al.* Nintendo Wii™ Versus Xbox Kinect™ for Assisting People With Parkinson's Disease. *Percept. Mot. Skills* **125**, 546–565 (2018).
39. Eftekharsadat, B. *et al.* Effect of virtual reality-based balance training in multiple sclerosis. *Neurol. Res.* **37**, 539–544 (2015).
40. Lee, N.-Y., Lee, D.-K. & Song, H.-S. *Effect of virtual reality dance exercise on the balance, activities of daily living, and depressive disorder status of Parkinson's disease patients*.
41. Adie, K. *et al.* Does the use of Nintendo Wii Sports™ improve arm function? Trial of Wii™

in Stroke: A randomized controlled trial and economics analysis. *Clin. Rehabil.* **31**, 173–185 (2017).
42. Bin Song, G., Cho Park, E., Song, G. Bin & Park, E. C. *Effect of virtual reality games on stroke patients' balance, gait, depression, and interpersonal relationships*. Journal of physical therapy science vol. 27 (2015).
43. Choi, H.-S. S., Shin, W.-S. S. & Bang, D.-H. H. Mirror Therapy Using Gesture Recognition for Upper Limb Function, Neck Discomfort, and Quality of Life After Chronic Stroke: A Single-Blind Randomized Controlled Trial. *Med. Sci. Monit.* **25**, 3271–3278 (2019).
44. Da Silva Ribeiro, N. M. *et al.* Virtual rehabilitation via Nintendo Wii(R) and conventional physical therapy effectively treat post-stroke hemiparetic patients. *Top. Stroke Rehabil.* **22**, 299–305 (2015).
45. Lee, H.-C., Huang, C.-L., Ho, S.-H. & Sung, W.-H. The Effect of a Virtual Reality Game Intervention on Balance for Patients with Stroke: A Randomized Controlled Trial. *Games Health J.* **6**, 303–311 (2017).
46. Henrique *et al.* Effects of Exergame on Patients' Balance and Upper Limb Motor Function after Stroke: A Randomized Controlled Trial. *J. Stroke Cerebrovasc. Dis.* **28**, 2351–2357 (2019).
47. Huang, L.-L. & Chen, M.-H. Effectiveness of the Immersive Virtual Reality in Upper Extremity Rehabilitation. in *Cross-Cultural Design. Applications in Health, Learning, Communication, and Creativity* (ed. Rau, P.-L. P.) 89–98 (Springer International Publishing, 2020).
48. Ikbali Afsar, S., Mirzayev, I., Umit Yemisci, O. & Cosar Saracgil, S. N. Virtual Reality in Upper Extremity Rehabilitation of Stroke Patients: A Randomized Controlled Trial. *J. Stroke Cerebrovasc. Dis.* **27**, 3473–3478 (2018).
49. In, T., Lee, K. & Song, C. Virtual Reality Reflection Therapy Improves Balance and Gait in Patients with Chronic Stroke: Randomized Controlled Trials. *Med. Sci. Monit.* **22**, 4046–4053 (2016).
50. Kiper, P. *et al.* Virtual Reality for Upper Limb Rehabilitation in Subacute and Chronic Stroke: A Randomized Controlled Trial. *Arch. Phys. Med. Rehabil.* **99**, 834-842.e4 (2018).
51. Llorens, R. *et al.* Improvement in balance using a virtual reality-based stepping exercise: a randomized controlled trial involving individuals with chronic stroke. *Clin. Rehabil.* **29**, 261–268 (2015).
52. Mekbib, D. B. *et al.* A novel fully immersive virtual reality environment for upper extremity rehabilitation in patients with stroke. *Ann. N. Y. Acad. Sci.* **1493**, 75–89 (2021).
53. Schuster-Amft, C. *et al.* Effect of a four-week virtual reality-based training versus conventional therapy on upper limb motor function after stroke: A multicenter parallel group randomized trial. *PLoS One* **13**, e0204455 (2018).
54. Shin, J. H. *et al.* Effects of virtual reality-based rehabilitation on distal upper extremity function and health-related quality of life: A single-blinded, randomized controlled trial. *J. Neuroeng. Rehabil.* **13**, (2016).
55. Zondervan, D. K. *et al.* Home-based hand rehabilitation after chronic stroke: Randomized, controlled single-blind trial comparing the music glove with a conventional exercise program. *J. Rehabil. Res. Dev.* **53**, 457–472 (2016).
56. Shih, M. C., Wang, R. Y., Cheng, S. J. & Yang, Y. R. Effects of a balance-based exergaming intervention using the Kinect sensor on posture stability in individuals with Parkinson's disease: A single-blinded randomized controlled trial. *J. Neuroeng. Rehabil.* **13**, (2016).
57. Kong, K.-H. H. *et al.* Efficacy of a Virtual Reality Commercial Gaming Device in Upper Limb Recovery after Stroke: A Randomized, Controlled Study. *Top. Stroke Rehabil.* **23**, 333–340 (2016).
58. Gandolfi, M. *et al.* Virtual Reality Telerehabilitation for Postural Instability in Parkinson's Disease: A Multicenter, Single-Blind, Randomized, Controlled Trial. *Biomed Res. Int.* **2017**, (2017).
59. Allen, N. E. *et al.* An interactive videogame for arm and hand exercise in people with

Parkinson's disease: A randomized controlled trial. *Parkinsonism Relat. Disord.* **41**, 66–72 (2017).
60. Ballester, B. R. *et al.* The visual amplification of goal-oriented movements counteracts acquired non-use in hemiparetic stroke patients. *J. Neuroeng. Rehabil.* **12**, 50 (2015).
61. Brunner, I. *et al.* Virtual Reality Training for Upper Extremity in Subacute Stroke (VIRTUES): A multicenter RCT. *Neurology* **89**, 2413–2421 (2017).
62. Maier, M., Rubio Ballester, B., Duff, A., Duarte Oller, E. & Verschure, P. F. M. J. Effect of Specific Over Nonspecific VR-Based Rehabilitation on Poststroke Motor Recovery: A Systematic Meta-analysis. *Neurorehabil. Neural Repair* **33**, 112–129 (2019).
63. Cameirao, M. S., Badia, S. B. i, Duarte, E., Frisoli, A. & Verschure, P. F. M. J. The combined impact of virtual reality neurorehabilitation and its interfaces on upper extremity functional recovery in patients with chronic stroke. *Stroke* **43**, 2720–2728 (2012).
64. Durgin, F. H. & Li, Z. Controlled interaction: strategies for using virtual reality to study perception. *Behav. Res. Methods* **42**, 414–420 (2010).
65. Piggott, L., Wagner, S. & Ziat, M. Haptic Neurorehabilitation and Virtual Reality for Upper Limb Paralysis: A Review. *Crit. Rev. Biomed. Eng.* **44**, 1–32 (2016).
66. You, S. H. *et al.* Virtual reality-induced cortical reorganization and associated locomotor recovery in chronic stroke: An experimenter-blind randomized study. *Stroke* **36**, 1166–1171 (2005).
67. Benham, S., Kang, M. & Grampurohit, N. Immersive Virtual Reality for the Management of Pain in Community-Dwelling Older Adults. *OTJR Occup. Particip. Heal.* **39**, 90–96 (2019).
68. Andersen, K. K., Olsen, T. S., Dehlendorff, C. & Kammersgaard, L. P. Hemorrhagic and ischemic strokes compared: stroke severity, mortality, and risk factors. *Stroke* **40**, 2068–2072 (2009).
69. Bhalla, A., Wang, Y., Rudd, A. & Wolfe, C. D. A. Differences in outcome and predictors between ischemic and intracerebral hemorrhage: the South London Stroke Register. *Stroke* **44**, 2174–2181 (2013).

*List of figures:*

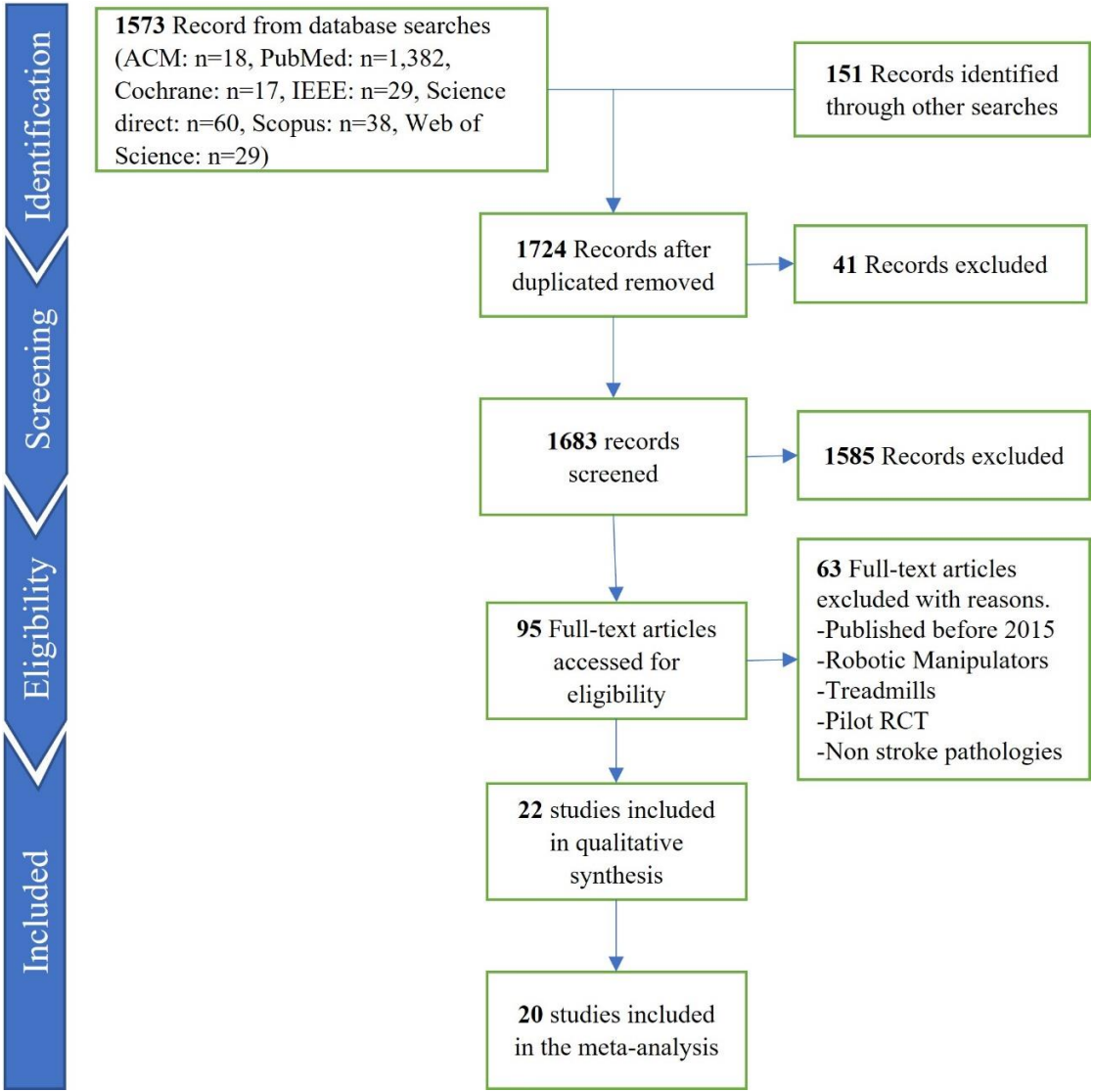

*Figure 1: Preferred reporting items for systematic reviews and meta-analyses flowchart of study retrieval, screening, and eligibility*

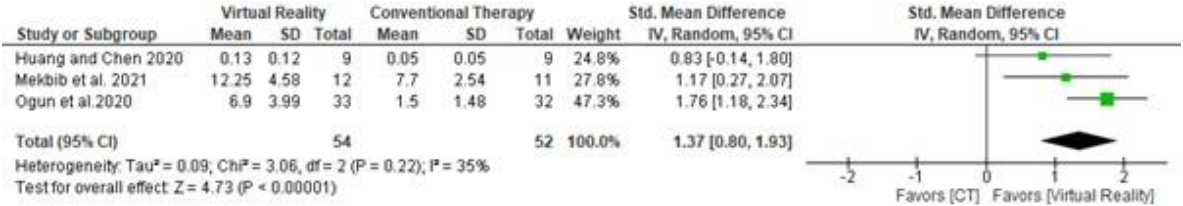

Figure 2: Forest plot of comparison: 1 IVR Versus CT (Short term Effects), outcome: FM

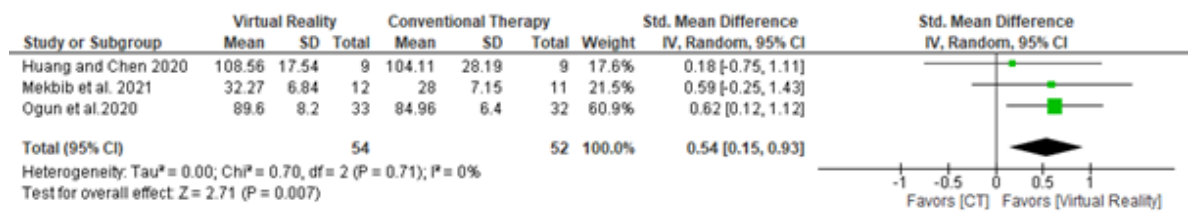

figure 3: Forest plot of comparison: IVR Versus CT (Short term Effects), outcome: Activities of Daily Life.

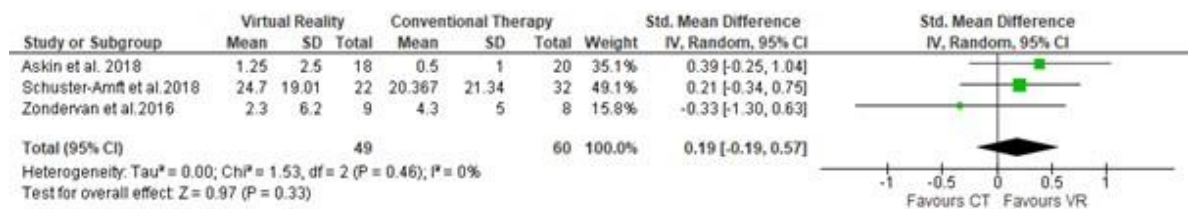

figure 4: Forest plot of comparison: NIVR Vs CT (Short term Effects), outcome: 3.3 BBT.

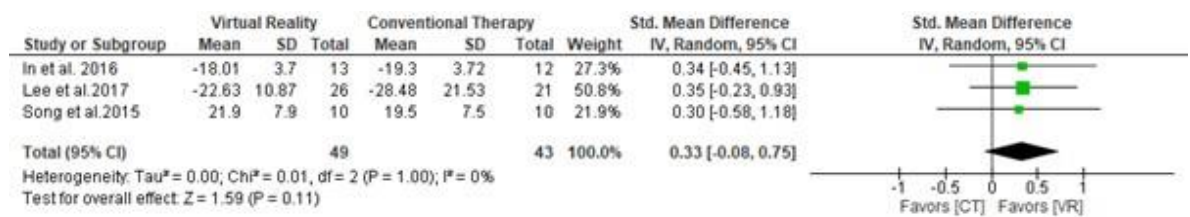

figure 5 Forest plot of comparison: Non-Immersive VR Vs CT (Short term Effects), outcome: TUG.

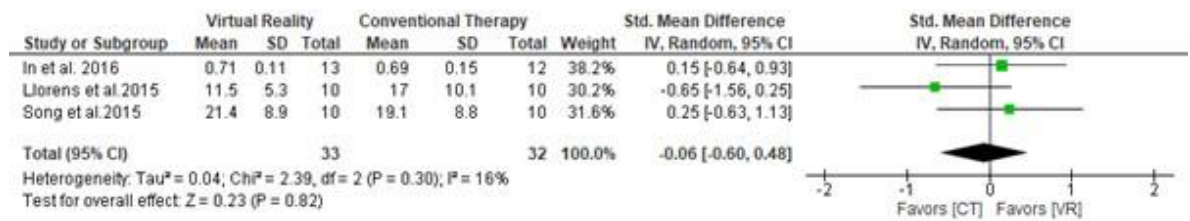

figure 6. Forest plot of comparison: NIVR Vs CT(Short term Effects), outcome: 10mW.

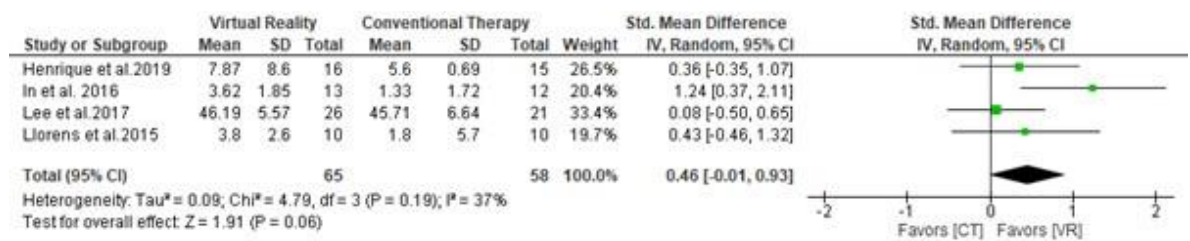

figure 7 Forest plot of comparison: NIVR Vs CT(Short term Effects), outcome: BBS.

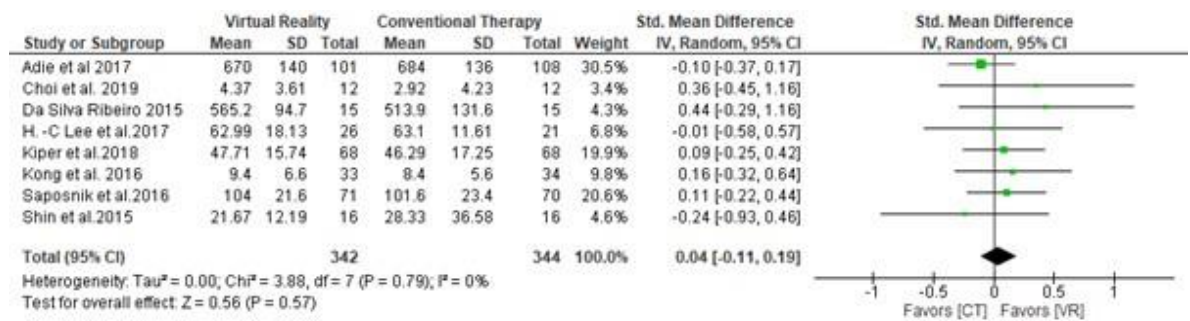

Figure 8: Forest plot of comparison: NIVR Vs CT(Short term Effects), outcome: QOL.

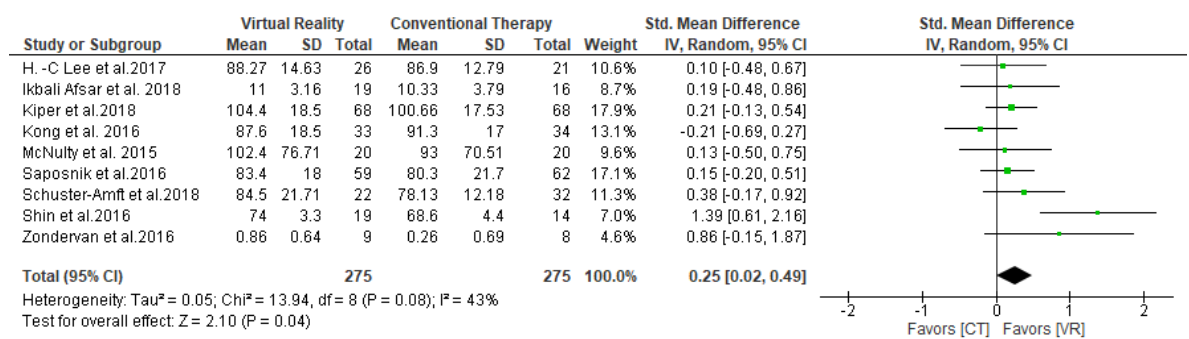

Figure 9: Forest plot of comparison: NIVR Vs CT(Short term Effects), outcome: ADL.

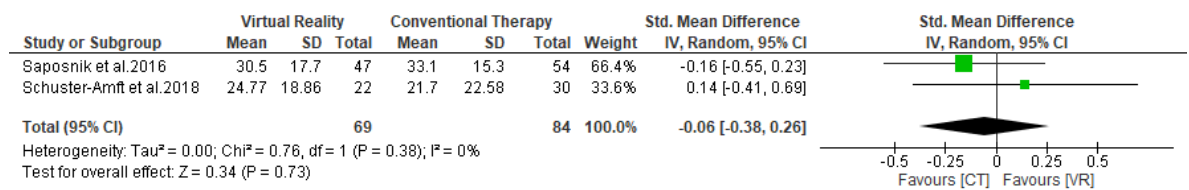

Figure 10: Forest plot of comparison: NIVR Vs CT (Long Term Effects), outcome: Upper Limb Activity. BBT.

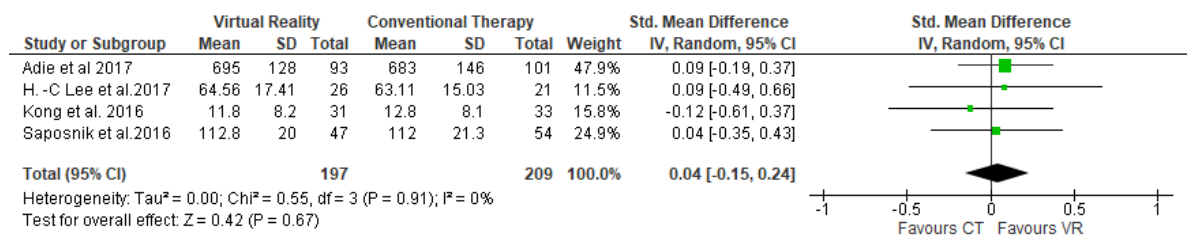

Figure 11: Forest plot of comparison: NIVR Vs CT (Long Term Effects), outcome: Quality Of Life.

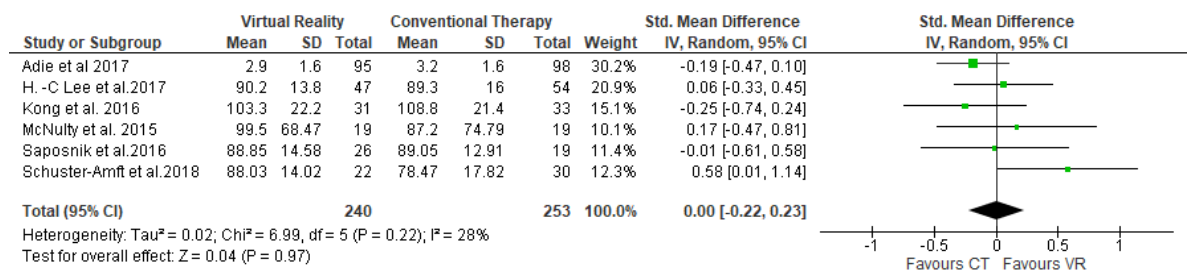

Figure 12: Forest plot of comparison: NIVR Vs CT (Long Term Effects), outcome: ADL.

*Table 1: Critical appraisal table. Indicates positive, negative responses from both authors, respectively shown in green, and red.*

| # | Study | 1. Did the trial address a focused issue? | 2. Was the assignment of patients to treatments randomized? | 3. Were all of the patients who entered the trial properly accounted for at its conclusion? | 4. Were patients, health workers, and study personnel blind to treatment? | 5. Were the groups similar at the start of the trial? | 6. Aside from the experimental intervention, were the groups treated equally? | 7. How large was the treatment effect? | 8. How precise was the estimate of the treatment effect? | 9. Can the results be applied to the local population, or in your context? | 10. Were all clinically important outcomes considered? | 11. Are the benefits worth the harms and costs? |
|---|---|---|---|---|---|---|---|---|---|---|---|---|
| 1. | Adie et al., 2017 [41] | ✓ | ✓ | ✓ | ✓ | ✓ | ✓ | ✗ | ✓ | ✓ | ✓ | ✓ |
| 2. | Aşkın et al. 2018 [23] | ✓ | ✓ | ✓ | ✓ | ✓ | ✓ | ✓ | ✗ | ✓ | ✓ | ✓ |
| 3. | Bin Song et al. 2015 [42] | ✓ | ✗ | ✗ | ✓ | ✗ | ✓ | ✓ | ✗ | ✓ | ✓ | ✓ |
| 4. | Choi, et al. 2019 [43] | ✓ | ✓ | ✓ | ✓ | ✓ | ✓ | ✗ | ✗ | ✓ | ✓ | ✓ |
| 5. | Da Silva Ribeiro et al. 2015 [44] | ✓ | ✓ | ✓ | ✓ | ✓ | ✓ | ✗ | ✗ | ✓ | ✓ | ✓ |
| 6. | H.-C. Lee et al. 2017 [45] | ✓ | ✓ | ✓ | ✓ | ✓ | ✓ | ✗ | ✗ | ✓ | ✓ | ✓ |
| 7. | Henrique et al. 2019 [46] | ✓ | ✓ | ✓ | ✓ | ✓ | ✓ | ✓ | ✓ | ✓ | ✓ | ✓ |
| 8. | Huang and Chen 2020 [47] | ✓ | ✗ | ✗ | ✓ | ✓ | ✓ | ✓ | ✗ | ✓ | ✓ | ✓ |
| 9. | Ikbali Afsar et al. 2018 [48] | ✓ | ✓ | ✓ | ✓ | ✓ | ✗ | ✓ | ✓ | ✓ | ✓ | ✓ |
| 10. | In, Lee, and Song 2016 [49] | ✓ | ✓ | ✓ | ✓ | ✓ | ✓ | ✓ | ✗ | ✓ | ✓ | ✓ |
| 11. | Kiper et al. 2018 [50] | ✓ | ✓ | ✓ | ✓ | ✓ | ✓ | ✗ | ✓ | ✓ | ✓ | ✓ |
| 12. | Kong et al. 2016 [27] | ✓ | ✓ | ✓ | ✓ | ✓ | ✓ | ✗ | ✓ | ✓ | ✓ | ✓ |
| 13. | Llorens et al. 2015 [51] | ✓ | ✓ | ✓ | ✓ | ✓ | ✓ | ✗ | ✗ | ✓ | ✓ | ✓ |
| 14. | McNulty et al., 2015, [30] | ✓ | ✓ | ✓ | ✓ | ✓ | ✓ | ✗ | ✗ | ✓ | ✓ | ✓ |
| 15. | Mekbib et al. 2021 [52] | ✓ | ✓ | ✓ | ✓ | ✓ | ✓ | ✓ | ✗ | ✓ | ✓ | ✓ |
| 16. | Ögün et al. 2019 [17] | ✓ | ✓ | ✓ | ✓ | ✓ | ✓ | ✓ | ✓ | ✓ | ✓ | ✓ |
| 17. | Pedreira da Fonseca et al. 2017 [29] | ✓ | ✓ | ✓ | ✓ | ✓ | ✓ | ✗ | ✗ | ✓ | ✓ | ✓ |
| 18. | Saposnik et al. 2016 [8] | ✓ | ✓ | ✓ | ✓ | ✓ | ✓ | ✗ | ✗ | ✓ | ✓ | ✓ |
| 19. | Schuster-Amft et al. 2018 [53] | ✓ | ✓ | ✓ | ✓ | ✓ | ✓ | ✗ | ✓ | ✓ | ✓ | ✓ |
| 20. | Shin et al. 2016 [54] | ✓ | ✓ | ✓ | ✓ | ✓ | ✓ | ✓ | ✗ | ✓ | ✓ | ✓ |
| 21. | Shin et al. 2015 [25] | ✓ | ✓ | ✓ | ✓ | ✓ | ✓ | ✗ | ✗ | ✓ | ✓ | ✓ |
| 22. | Zondervan et al. 2016 [55] | ✓ | ✓ | ✓ | ✓ | ✓ | ✓ | ✗ | ✗ | ✓ | ✓ | ✓ |

Table 2: IVR vs CT Participant Information.

| | **Study** | **Study Type** | **Participants** | | | |
|---|---|---|---|---|---|---|
| | | | population | pathology duration (mean ± SD), | age (years) (Mean ± SD) | Sex |
| 1. | Huang and Chen 2020[47] | SC | n=17,VR(9),CT(9) | CT(7.87±7.07),IVR:(9.69±3.5) | ,IVR(59.48±15.02),CT (55.36±10.48) | CT(F:2,M:7),IVR(F:1,M:8) |
| 2. | Mekbib et al. 2021[52] | SC | n=23,VR(12),CT(11) | IVR(1.23±0.73),CT(39.36±18.08) | IVR(52.17±13.26),CT (61.00 ± 7.69) | IVR:(M:9,F:3),CT(M:8,F:3) |
| 3. | Ögün et al. 2019[17] | SC | n=84,VR(42),CT(42) | IVR:(1.32±0.25),CT:(0.51±0.32) | IVR:(61.5±10.9) CT:(59.8 ± 8.1) | IVR:(M:28,F:5)CT:(M:23,F:9) |

Abbreviations: IVR: Immersive Virtual Reality; C: control, CT: Conventional Therapy, mo: months

Table 3: NIVR vs C Participant Information.

| | Study | Study type | Participants | | | |
|---|---|---|---|---|---|---|
| | | | Population | pathology duration. (mo) (mean ± SD) | age (yrs) (mean ± SD) | Sex |
| 1. | Adie et al. 2017[41] | SC | n=235,VR(117),CT(118) | VR(1.91±1.61),CT(1.88 ±1.67) | VR(66.8±14.6),CT (68.0±11.9) | VR(F:51,M:66),CT (F:53, M:65) |
| 2. | Aşkın et al. 2018[23] | SC | n=40,VR(20),CT(20) | VR(20.27±5.47),CT(19.40±4.48) | VR(53.3±11.2), CT(56.6 ± 9.9) | VR(F:5,M:13),CT(F:6,M:14) |
| 3. | Bin Song et al. 2015[42] | SC | n=40,VR(20),CT(20). | VR(14.8±6.1), CT(14.3±3.4) | VR(51.37±40.6) ,CT(50.10±7.83) | VR(M:10,F:10),CT(M:12,F:8) |
| 4. | Choi, Shin, and Bang 2019[43] | SC | n=36,VR(12),CT(12),C(12) | CT(28.91±15.80), VR(26.33±15.51),C(29.00±19.21) | CT(58.00±5.15) ,VR(59.58±11.87), (59.33±13.63) | CT:(M:7,F:5),VR(M:7,F:5),C(M:9, F:3) |
| 5. | Da Silva Ribeiro et al. 2015[44] | SC | n=30,VR(15), CT(15) | VR(42.1±26.9),CT(60.4±44.1), | VR(53.7±6.1), CT(52.8±8.6) | VR(M:06,F:09),CT(M:05,F:10) |
| 6. | H.-C. Lee et al. 2017[45] | SC | N=50,VR(26), CT(24) | VR(28.00±23.97), CT(21.77 ± 19.66) | VR(59.4±8.95), CT(55.8 ± 9.6) | VR(M:16,F:10), CT(M:18, F:3 ) |
| 7. | Henrique et al. 2019[46] | SC | n=31,VR(16) CT(15) | CT(17.07 ± 10.00),VR (15.63 ± 6.60) | CT(76.20±10.41),VR( 76.19±10.09) | CT(M:7,F:8),VR (M:7,F:9) |
| 8. | Ikbali Afsar et al. 2018[48] | SC | n = 35,VR(19) , CT(16) | VR(2.94±1.88),CT(2.29 ±1.31) | VR(69.4±8.6),CT (63.4 ± 15.7) | VR(F:7,M:12,CT( F:8, M:8) |
| 9. | In, Lee, and Song 2016[49] | SC | n=25,VR(13), CT (12) | VR(12.5±4.1), CT(13.6±5.3) | VR(57.3±10.5),CT(54.4±11.4) | VR(M:8,F:5),CT (M:7, F:5) |
| 10. | Kiper et al. 2018[50] | SC | (N=136,VR(68), CT(68) | VR(52.8.4±33.6),CT(49.2±38.4) | VR(62.5±15.2), CT (66.0±12.9) | VR(M:37,F:31) , CT(M:43, F:25) |
| 11. | Kong et al. 2016[27] | SC | N=105,VR(35),CT (35), C: (35) | VR(0.47±0.30),CT(0.47 ± 0.31), C(0.44 ± 0.29) | VR(58.1±9.1),CT(59.0 ±13.6),C(55.8 ± 11.5) | VR(M:27,F:9), CT(M:25,F:10) C(M:25 F:10) |
| 12. | Llorens et al. 2015[51] | SC | n=20,CT(10), VR (10) | CT(19.59±7.4),VR (13.58 ±7.75) | CT(55.0±11.6),VR (58.3 ±11.6) | CT(M:5,F:5), VR(M:4, F:6) |
| 13. | McNulty et al. 2015[30] | SC | n=41,VR(21) CT (20) | VR(11·0±3·1),CT(6·5± 1) | VR(59·9±13·8) CT(56·1 ± 17·0) | VR(F:8,M:13), CT(F:2,M:18) |
| 14. | Pedreira da Fonseca et al. 2017[29] | SC | n=27,VR(14), CT(13) | VR(44.1±25.0),CT(64.5 ±41.9) | VR(53.8±6.3), CT(50.9±10.9) | VR(F:10,M:4),CT(F:9,M:4) |
| 15. | Saposnik et al.2016[8] | MC | n=141,VR(71), CT(70) | VR(102±16·8)CT(102 ±19·2) | VR(62±13)CT(62 ± 12) | VR(M:46,F:25), CT(M:48, F:31) |
| 16. | Schuster-Amft et al. 2014[24] | MC | n=42,VR(22),CT(32) | VR(28.8±28.8),CT(43.2±44.4) | VR(61.3±13.4), (CT 61.2±11.2) | VR(F:6,M:16),CT (F:9, M: 23) |
| 17. | Shin et al. 2016[54] | SC | n=46,VR(24), CT(22) | VR(13.6±13.4),CT(15.0 ±14.6) | VR(57.2±10.3), CT(59.8 ± 13.0) | VR(M:19,F:5),CT (M:17, F:5) |
| 18. | Shin, et al. 2015[25] | SC | n=32,VR(16),CT(16) | VR(6.73±2.96), CT(5.5±2.91) | VR(53.37±11.8), CT( 54.67± 13.4) | VR(M:11,F:5), CT(M:13,F: 3) |
| 19. | Zondervan et al. 2016[55] | SC | n=17,VR(9), CT(8) | VR(63.96 ± 49.68), CT(38.04±19.92) | VR(59.78±9.71) CT(59.95 ±13.60) | VR(F:4,M:5), CT(F:3,M:5) |

Abbreviations: VR-Virtual Reality; C-control, CT-Conventional Therapy, SC-Single Center, MC-Multi Center, yrs-years, mo - months

Table 4: IVR Interventions, Comparator, and Technology

| | Study | Intervention | Comparator | VR Technology |
|---|---|---|---|---|
| 1. | Huang and Chen 2020 [47] | 30 mins IVR games + 60 mins CT including a Climbing bar, Ball bearing, and Pulley for 20 sessions. Tot. 30 hrs | 90 mins CT of Upper limb training using a Climbing bar, Ball bearing, and Pulley for 20 sessions. Tot. 30 hrs. | HTC VIVE HMD, Controllers |
| 2. | Mekbib et al. 2021 [52] | 60min IVR activities like reaching, grasping, and releasing tasks + 60min CT for 8 sessions. Tot 16 hrs. | 120 min CT of daily living activities, balance control, gait training, weight shift, and distal and proximal UE functional movements for 8 sessions, Tot 16 hrs. | HTC VIVE HMD + Leap Motion |
| 3. | Ögün et al. 2019 [17] | 60 min IVR activities to facilitate hand motions, stimulating forearm supination and pronation, flexion, and abduction 18 sessions. Tot: 18 hrs. | 60 min CT of upper extremity exercises comprising the same tasks as used in the IVR group.+ 15min Sham IVR for 18 sessions. Tot:18 hrs. | Leap Motion, HMD |

Abbreviations: IVR-Immersive Virtual Reality, CT-Conventional Therapy, HMD-Head Mounted display

Table 5: NIVR Interventions, Comparator, Technology.

| | study | Intervention | Comparator | VR Technology |
|---|---|---|---|---|
| 1. | Adie et al. 2017[41] | 15 min warm-up + 45 min Wii VR per day; Tot: 17hrs | 15 min warm-up + 45 min CT per day. Tot: 17 hrs | Nintendo Wii |
| 2. | Aşkın et al. 2018[23] | 20 sessions of CT + VR games that required the upper extremity use, Tot 20 hrs. | 20 sessions of CT activities to improve the active range of motion, strength, flexibility, transfers, posture, balance, coordination, and activities of daily living. Tot 20 hrs. | Xbox Kinect, TV screen, laptop |
| 3. | Bin Song et al. 2015[42] | VR 30 min/session, Tot 20 hrs (Kinect games for body balance and limb motion) | Ergometer bicycle training 30 min/session, Tot 20 hrs | Xbox Kinect |
| 4. | Choi, Shin, and Bang 2019[43] | 30 min CT + VR mirror therapy 30 min of lifting the arms, moving the arms to the left and right, bending and stretching the elbows, raising and lowering the hands, lifting the wrists, lowering the wrists, flexing the wrists inward, flexing the wrist, and finger gripping,15 sessions. Tot 15 hrs. | 30 min CT + 30min Conventional Mirror Therapy: 15 sessions. Tot 15 hrs. | Leap motion, a monitor, a mirror |
| 5. | Da Silva Ribeiro et al. 2015[44] | 10 min stretching of UL, LL, and trunk muscles. + 50-min of VR games. 8 sessions Tot 16 hrs. | 10-min stretching + 50min CT of trunk activities, active or active-assisted diagonal movement of the Lower Limbs, balance training, stationary and side gait, anteroposterior and laterolateral movements, gait training. 8 sessions. Tot 16 hrs. | Nintendo Wii, projector |
| 6. | H.-C. Lee et al. 2017[45] | 45 min CT + 45 min VR balance games based on common balance problems experienced after stroke. for 12 sessions Tot 18 hrs | CT for 90 min focusing on strengthening, endurance training, ambulation, and ADL training. for 12 sessions. Tot 18 hrs. | television, Microsoft Kinect + commercial game |
| 7. | Henrique et al. 2019[46] | VR exergame 30 mins for 24 sessions weeks, Tot 12 hrs. exergame for upper limb motor function and balance rehabilitation of stroke survivors, including flexion exercises, shoulder abduction and adduction, horizontal shoulder abduction and adduction, elbow extension, wrist extension, knee flexion, hip flexion, and abduction | CT 30 mins exercise similar to VR group, for 24 sessions, Tot 12 hrs. | Motion Rehab 3D, projector, Kinect, PC |
| 8. | Ikbali Afsar et al. 2018[48] | CT 60 min + VR 30 min programs for active movements of the upper extremity, bilateral shoulder abduction, and adduction, and active elbow flexion and extension movements, performed flexion and extension movements in both the shoulder and elbow joints. For 20 sessions. Tot 30 hrs. | CT 60 min consisting of static and dynamic control of position, balance skills, weight shift, and activities of daily living for 20 sessions. Tot 20 hrs. | Microsoft Xbox Kinect system, TV screen |
| 9. | In, Lee, and Song 2016[49] | CT 30 min + VR 30 min, for 20 sessions. Tot 20 hrs. Participants placed their affected lower limb into the VRRT box to observe the projected movement of the unaffected lower limb without visual asymmetry causing tilting of the head and trunk. | CT 30 min of neurodevelopmental treatment, physical therapy, occupational therapy, and speech therapy. + placebo VR 30 min, for 20 sessions. Tot 20 hrs, consists | camcorder, LCD monitor |
| 10. | Kiper et al. 2018[50] | VR 1hr tasks which consisted of both simple movements and complex movements that involved multiple muscle synergies + CR 1hr for 20 sessions: Tot 40 hrs. Virtual | CT 2 hrs of upper limb exercises such as shoulder flexion and extension, shoulder abduction and adduction, shoulder internal and external rotation, elbow flexion and extension, forearm pronation and supination and hand grasping-release tasks for 20 sessions: Tot 40 hrs. | 3-D motion tracking system, projector |
| 11. | Kong et al. 2016[27] | 1hr VR games for executing movements and acceleration of the upper limbs 4 times/ wk. over 3 wks, plus + 1 hour of PT from Mon to Friday. Tot 27 hrs. | 1 hr CT of stretching, strengthening, and upper limb range of motion exercises.4 times/wk. for 3 wks + 1 hr of PT from Mon to Fri. Tot 27 hrs. | Nintendo Wii |
| 12. | Llorens et al. 2015[51] | 30min VR games for a stepping task, + 30min CT. 20 sessions. Tot 20 hrs. Games | 1 hr CT of static standing exercises, task-specific reaching exercises involving ankle and hip, stepping tasks, static and dynamic balance exercises, walking exercises for 20 sessions. Tot 20 hrs. | PC, Screen, and an optical tracking system |
| 13. | McNulty et al. 2015[30] | VR, 60-min of Wii Sports games for the more affected hand + home practice. Tot 10 hrs. | mCIMT 60-min of Training tasks including everyday activities using only the more affected hand and arm + home practice. Tot 10 hrs. | Nintendo Wii |
| 14. | Pedreira da Fonseca et al. 2017[29] | 15min stretch +45 min VR games which stimulated the lateralization of movements of the trunk; weight shift between the heel and forefoot, working rotational movements of the trunk, weight transfer between the heel and forefoot, rotational movements of the hip, and balance reaction time. for 20 sessions Tot 20 hrs. | 10min stretch for arm and leg muscles + 50 min CT trunk mobilization activities in the lateral, anterior, and posterior directions(10min), leg movement;(15min) balance training in a standing position (10 min); and free gait training for 10 mins. for 20 sessions Tot 20 hrs. | Nintendo Wii, projector |
| 15. | Saposnik et al. 2016[8] | VR 60 min with the goals of enhancing flexibility, range of motion, strength, and coordination of the affected arm 10 sessions, Tot 10 hrs. | recreational therapy, 10 sessions, 60 min Tot 10 hrs. | Nintendo Wii |
| 16. | Schuster-Amft et al. 2014[24] | VR 45-min for use of the arm and/or hand movements, mirroring of the real movements of one arm and/or hand and following the movements of one arm and/or hand. ,for 16 sessions Tot: 12hrs. | 45-min CT which included neuromuscular interventions, body structural interventions perceptual and sensory interventions, 16 sessions. Tot:12hrs. | PC, gloves. |
| 17. | Shin et al. 2016[54] | VR 30min + 30min OT for movements of the distal upper extremity such as the forearm pronation/supination, wrist flexion/extension, wrist radial/ulnar deviation, finger flexion/extension, and complex movements .20 sessions tot: 20hrs. | 60 min OT (20 sessions) Tot: 20hrs. Same categories of movements of the distal upper extremity as those in the VR group | smart glove |
| 18. | Shin, Bog Park, and Ho Jang 2015[25] | 30 min CT + 30 min VR games for active arm and trunk movements and promote successful rehabilitation. 20 sessions. Tot 20 hrs. | 1 hr of CT which includes a range of motion and strengthening exercises for the affected limb, table-top activities, and training for activities of daily living for 20 sessions. Tot 20 hrs. | Depth sensor, 3D awareness sensors, infrared projectors, and image sensors. |

| 19. | Zondervan et al. 2016[55] | 1hr VR games of self-guided therapy for hand and finger exercises for 9 sessions, Tot: 9hrs | 1hr CT of self-guided therapy of tabletop hand and finger exercises 1hr for 9 sessions, Tot: 9hrs | Music Glove, laptop |

Abbreviations: VR Virtual Reality; C control, CT Conventional Therapy, hr hour

Table 6. Study Outcomes of IVR Vs. CT

|   | Study | Upper limb function | Activity limitation/ ADL | Results |
|---|---|---|---|---|
| 1. | Huang and Chen 2020[47] | Pri:FM(IVR:0.13±0.12,CT:0.05±0.05,ES:,0.83), BBT | FIM | Greater improvement in VR in FM and FIM |
| 2. | Mekbib et al. 2021[52] | Pri:FM(IVR:12.25±4.58,CT:7.7±2.54, ES:,1.17), | BI | Greater improvement in VR on FM |
| 3. | Ögün et al. 2019[17] | Pri:FM(IVR:46.54±7.91,CT:40.06±8.33,ES:0.79,) ARAT | PASS-BADL, PASS-IADL, FIM | Greater improvement in VR on FM, ARAT, FIM and PASS |

Table 7. Study Outcomes for NIVR Vs CT

| | Study | Outcomes | | | | | | Results |
|---|---|---|---|---|---|---|---|---|
| | | Upper limb function | Lower limb function | Balance and postural control | QOL | Activity limitation (ADL) | Adverse events | |
| 1. | Adie et al. 2017[41] | Pri: ARAT (NIVR: 47.6±14.2, CT:49±13.6, ES:0.1), | | | SIS, COPMS, COPMP | MAL-QOM, MAL -AOU | | no significant difference between groups in all outcomes. |
| 2. | Aşkın et al. 2018[23] | Pri: FM(NIVR: 4.33±7.24, CT:0.67±1.61,ES: 0.64),BBT, AROM, BRS, MAS Hand | | | | | Number | Greater improvements in FM in VR. |
| 3. | Bin Song et al. 2015[42] | | TUG(NIVR:21.9±7.9,CT:19.5±7.5,ES:0.31),10mWT(NIVR:214.2±8.9,CT:19.1±8.8.ES:0.26) | balance ability, (NIVR:24.7±19.01,CT:20.37±21.34,ES:0.61)forward LOS,(NIVR:3311.7±19.01,CT:4322.6± 565.5 ES:0.28)Backward LOS (NIVR:1895.9±2097.5,CT:2889.7±2769.7,ES:0.4) | | | | Greater improvements in VR in weight distribution ratio, anterior LOS, posterior LOS, TUG, 10-mWT, and BDI |
| 4. | Choi, Shin, and Bang 2019[43] | Pri:MFT(NIVR:13.42±2.5,CT:12.33±2.02 ,ES:0.46), | | | SF- 8 | | | Greater improvement in FM in VR. |
| 5. | Da Silva Ribeiro et al. 2015[44] | Pri: FM(NIVR: 46.54±7.91,CT:40.06± 8.33, ES:0.34), | | | SF-36 | | | No inter-group difference in FM, greater improvement in VR on SF-36 |
| 6. | H.-C. Lee et al. 2017[45] | | TUGcog | Pri:BBS(NIVR:46.19±5.57, CT:45.7±6.64,ES:0.08), FRT | SIS | MBI, ABC scale , M-PAES | Number | No significant intergroup difference |
| 7. | Henrique et al. 2019[46] | Pri:FM(NIVR: 14.69±0.67, CT:9.07±1.34, ES:5.22), | | Pri:BBS(NIVR:7.87±8.6, CT:5.6±0.69: ES:0.25), | | | | Greater improvement in VR on FM. |
| 8. | Ikbali Afsar et al. 2018[48] | Pri:FM(NIVR:18.74±7.67,CT:13.94± 6.58,ES:0.65), BBT,BRS-arm,BRS-Hand, | | | | FIM | | Greater improvement in VR for BRS and BBT. |
| 9. | In, Lee, and Song 2016[49] | | TUG, 10-mwv | Pri:BBS(NIVR:3.62±1.85,CT:1.33±1.72.ES:1.24),FRT, postural sway, | | | | Greater improvement in VR for FRT, TUG, and 10 mWV. |
| 10. | Kiper et al. 2018[50] | Pri:FM(NIVR:47.1±15.7,CT:46.29±17.25, ES:0.09), | | | NIHSS, ESAS | FIM, | | Greater improvements in VR for FM, FIM, NIHSS |
| 11. | Kong et al. 2016[27] | Pri:FM,(NIVR:32.8±18.2,CT:29.2±1 | | | SIS | FIM | Number | no significant difference in all |

| # | Study | | | | | | | Outcome |
|---|---|---|---|---|---|---|---|---|
| | | | | 7.25,ES:0.2), ARAT | | | | outcome measures between groups. |
| 12. | Llorens et al. 2015[51] | | 10-mWTest. TP-OMA | Pri:BBS(NIVR:3.8±2.6,CT:1.8±5.7,ES:,0.43), BBA | | | | greater improvement VR for BBS and 10mWT. |
| 13. | McNulty et al. 2015[30] | WMFT, FMA | | | | Pri:MALQoM(NIVR:102.±38.4,CT:93.±35.3,ES:0.25) | Number | No differences between groups for WMFT and MALQoM |
| 14. | Pedreira da Fonseca et al. 2017[29] | | | Pri:DGI,(NIVR:-.71±3.14,CT:-2.84±4.63,ES:0.27), | | | No of falls | no significant difference in DGI and number of falls. |
| 15. | Saposnik et al. 2016[8] | Pri:WMFT, (NIVR:-64.1±104, CT:-39.8±35.5, ES:0.31),BBT, Grip Strength | | | SIS | BI, FIM, | Number | no significant difference between groups in WMFT. |
| 16. | Schuster-Amft et al.2014[24] | Pri:BBT(NIVR:24.7±19.01,CT:20.37±21.34,ES:,0.21) | | | | CAHAI | Number | no between-group differences for all outcomes. |
| 17. | Shin et al. 2016[54] | Pri:FM(NIVR:4.0±1.0,CT:1.4±0.8,ES:3.71), JTT, PGT | | | | SIS | Number | Significant improvements in the FM and SIS in VR group during the intervention and at follow-up; |
| 18. | Shin, et al. 2015[25] | Pri:FM(NIVR:38.5±11.7,CT:33.87±17.64, ES:0.3), | | | SF-36 | | | No inter-group differences in FM. Greater improvement in VR for role limitation due to physical problems. |
| 19. | Zondervan et al. 2016[55] | Pri:BBT(NIVR:2.3±6.2,CT:4.3±5,ES:0.33) NHPT,ARAT | | | | MALQOM, MAL (AOU) | Number, pain | no significant difference between groups for BBT. VR had significantly greater improvements in MALQoM and AoU 1 mo posttherapy |

10mWT-10m Walking Test; ADL-Activities of Daily Life; ABC-Activities-specific Balance Confidence; AOU-Amount of use; ARAT-Action Research Arm test; AROM-active range of motion; BBA-Brunel Balance Assessment; BBS-Berg Balance Scale; BBT-Box and Blocks Test; BDI-Beck Depression Inventory; BI- Barthel Index; BRS-Brunnstrom Recovery Stages; C-Control; CT-Conventional Therapy; CAHAI -Chedoke-McMaster Arm and Hand Activity Inventory; COPMP-Canadian Occupational Performance Measure Performance; COPMS-Canadian Occupational Performance Measure Satisfaction; DGI-Dynamic Gait Index; ES-Effect Size: ESAS -Edmonton Symptom Assessment Scale; FM-Fugl-Meyer; FIM-Functional Independence Measure ; FRT-Functional Reach Test; IVR-Immersive Virtual Reality; JTT -Jebsen–Taylor hand function test; LOS-Limits of stability; MAL-Motor Activity Log; MAS-Modified Ashworth Scale; MBI-Modified Barthel Index; MFT-Manual Function test; M-PAES-Modified Physical Activity Enjoyment Scale; NHPT-Nine Hole Peg Test; NIHSS-National Institutes of Health Stroke Scale; NIVR-Non- Immersive Virtual reality; PASS-BADL-Performance Assessment of Self-Care Skills – basic activities of daily living; PASS-IADL-Performance Assessment of Self Care Skills – instrumental activities of daily living; PGT-Purdue pegboard test; QoM-Quality of Movement; SF-Short Form Health Survey; WMFT-Wolf Motor Function Test